\documentclass[a4paper]{revtex4}
\usepackage{graphicx}
\usepackage{fancyhdr}
\usepackage{amsmath}
\newcommand{\nl}{\nonumber\\}
\renewcommand{\ol}{\overline}
\newcommand{\bea}{\begin{eqnarray}}
\newcommand{\eea}{\end{eqnarray}}

\newcommand\fverbdo{\egroup\medskip\noindent%
			\fbox{\unhbox\fverbbox}\ }
\newcommand\fverbit{\egroup\item[\fbox{\unhbox\fverbbox}]}
\newbox\fverbbox

\newcommand{\be}{\beta}
\newcommand{\ga}{\gamma}
\newcommand{\la}{\lambda}
\newcommand{\si}{\sigma}
\renewcommand{\th}{\theta}

\newcommand{\De}{\Delta}

\usepackage{color}

\begin{document}

\title{Radiative neutrino mass and 3.5 keV X-ray line}

%

\author{Seungwon Baek}
\affiliation{School of Physics and Open KIAS Center, KIAS,\\ 85 Hoegiro Dongdaemun-gu, Seoul 130-722, Korea}

\begin{abstract}
We consider an extension of Zee-Babu model to explain the smallness of neutrino masses.
(1) We extend the lepton number symmetry of the original model to local $B-L$ symmetry.
(2) We introduce three  Dirac dark matter candidates with flavor-dependent $B-L$ charges.
After the spontaneous breaking of $B-L$, a discrete symmetry $Z_6$ remains, which guarantees the stability of dark matter.
Then the model can explain the 3.5 keV X-ray line signal with decaying dark matter.
We also introduce a real scalar field which is singlet under both the SM and $U(1)_{B-L}$ and can explain the current 
relic abundance of the Dirac fermionic DMs. If the mixing with the SM Higgs boson is small, 
it does not contribute to DM direct detection. The main contribution to the scattering of DM off atomic nuclei comes from 
the exchange of $U(1)_{B-L}$ gauge boson, $Z'$, and is suppressed below current experimental bound when $Z'$ mass is
heavy  ($\gtrsim 10$ TeV). If the singlet scalar mass is about 0.1--10 MeV, DM self-interaction can be large enough to solve small 
scale structure problems in simulations with the cold DM, such as, the core-vs-cusp problem and too-big-to-fail problem.
\end{abstract}

\maketitle

\thispagestyle{fancy}


\section{Introduction}
Although the existence of dark matter (DM) is now well-established from various observations, such as
rotation curves,
structure formation, elliptical galaxy NGC 720, gravitational lensing, bullet cluster, temperature fluctuation of 
cosmic microwave background,  {\it etc.}.
However, all of the above evidences result from the gravitational interactions of DM(s), and
the nature of DM(s) is still not well-known.
Thus the search for DM interactions, especially non-gravitational ones, is one of the hot topics in theoretical and
experimental physics.
The recently observed 3.5 keV X-ray line signal in a stacked spectrum of galaxies
and clusters of galaxies~\cite{Xray_exp}, if it is confirmed, can be a first strong  hint for the non-gravitational 
DM interaction. The conventional scenario for the X-ray line in terms of DM models 
is the decay of sterile neutrino with mass $m_s=7.06 \pm 0.5$ keV into a 3.5 keV photon and an active neutrinos.
The observed flux~\cite{Xray_exp} 
\bea
\Phi_{\rm X-ray} \propto n_s \Gamma_s &=& 1.39 \times 10^{-22}\, {\rm s}^{-1} \sin^2 2\th \left(m_s \over {\rm keV}\right)^5 
\frac{\rho_{\rm DM}}{m_s} \nl
&=& (1.5 \times 10^{-25} - 2.7 \times10^{-24})\, {\rm cm^{-3} s^{-1}},
\label{eq:flux}
\eea
can be explained by mixing angle given by $\sin^2 2\th =(2-20) \times 10^{-11}$.

It would be interesting to consider a model with an interplay between DM and other sectors of the SM, for example, 
the neutrino sector.
Then measurement of one sector may predict or constrain the other sector.
One of these scenarios has been studied in~\cite{Baek:2012ub}.
In \cite{Baek:2012ub}, we introduced scalar dark matter coupled to the Zee-Babu model
which generates neutrino masses radiatively at two-loop level~\cite{Zee-Babu}. 
We showed that the model can successfully explain Fermi-LAT 130 GeV gamma-ray line.

In this talk based on~\cite{Baek:2014}, we gauge the global $U(1)_{B-L}$ symmetry of \cite{Baek:2012ub}.
To cancel the gauge anomaly we need to introduce three right-handed neutrinos $N_{R_i} (i=1,2,3)$ with $B-L=-1$.
We also introduce a complex scalar field $\varphi$ with $B-L=2$ which breaks the local $U(1)_{B-L}$ symmetry
when $\varphi$ gets vacuum expectation value (VEV), $v_\varphi$.
The $\varphi$ field also generates the soft lepton number breaking term of original Zee-Babu model
dynamically.
The $U(1)_{B-L}$ symmetry allows the Yukawa interaction $\ell H N_{R_i}$ and Majorana mass terms $ N_{R_i} N_{R_i} \varphi$,
which will generate neutrino masses  through the usual seesaw mechanism after $U(1)_{B-L}$ symmetry is broken.
Since we want to generate the neutrino masses only through Zee-Babu mechanism~\cite{Zee-Babu},
we forbid the above Yukawa interaction by introducing a global $Z_2$ symmetry
under which only $N_{R_i}$ are odd
and all other particles are even.

We introduce Dirac fermionic dark matter candidates $\psi_i (i=1,2,3)$  to explain the X-ray line signal.
The $\psi_i$ are neutral under the SM gauge group
but charged under the local $U(1)_{B-L}$ symmetry. They are vector-like under $U(1)_{B-L}$ symmetry
and gauge anomaly is not generated.
If we assign the $U(1)_{B-L}$ charges of $\psi_i$ fields in such a way that
$\Delta Q_{\psi} \equiv Q_{\psi_2} - Q_{\psi_1} = Q_{\psi_3} -Q_{\psi_2}=2$, {\it off-diagonal} Yukawa interactions, $\ol{\psi_1} \psi_2 \varphi^*$
and $\ol{\psi_2} \psi_3 \varphi^*$, are allowed. After $\varphi$ gets VEV,
off-diagonal terms in the mass matrix of $\psi$'s are generated, which induces the dark-flavor-changing $Z'$ couplings 
at tree level. And flavor-changing radiative decay of DM is allowed.
We can also see that a discrete symmetry remains after $U(1)_{B-L}$ symmetry is broken. 
This {\it local} discrete symmetry guarantees absolute stability of the lightest state of $\psi_i$~\cite{local_DM},
as opposed to the global symmetry which can be broken by quantum gravity.
The quantum gravity effect can break the global $Z_2$ symmetry, and the right-handed neutrinos can decay very fast
without causing cosmological problems such as BBN. The light singlet particle also decays before BBN.

We show that transition magnetic dipole operator (TMDO) $\ol{\psi'_1} \si_{\mu\nu} \psi'_2 F^{\mu\nu}/\Lambda$ 
($\psi'_i$ are mass eigenstates)
can be generated by two-loop diagrams involving Zee-Babu scalars, $\varphi$ scalar, and $B-L$ gauge boson.
The heavier state $\psi'_2$  can decay into the lighter state and photon through this TMDO.
If the mass difference between the two states is about $3.5$ keV, we can explain the observed
X-ray line signal. Since the TMDO is generated at two-loop level, the effective cut-off scale $\Lambda$ of the operator can be
very high, even if all the particles running inside the loop have (sub-)electroweak scale masses. 
As a consequence $\psi'_2$ can live much longer than the age of the universe and can be a decaying DM candidate.

{
In our model there appear some small parameters, such as
$v_\eta/v_\varphi$, $\Delta m_{21}/m_\psi$, {\it etc.}, 
which seems to be fine-tuning at first sight. However, we will show that they
are technically natural in the sense of 't Hooft:
\bea
\text {\it ``A parameter is naturally small if setting it to
  zero increases the symmetry of the theory.''}.
\label{itm:tHooft}
\eea
}

\section{The model and 3.5 keV line signal}
\label{sec:model}
The model contains two electrically charged Zee-Babu scalar fields 
$h^+$, $k^{++}$, a SM-singlet complex  
dark scalar $\varphi$, a singlet real scalar $\eta$, three right-handed neutrinos $N_{R_i} (i=1,2,3)$
and  three SM-singlet Dirac fermion dark matter candidates $\psi_i$ in addition to
the SM fields. 
In Table~\ref{tab:B-L}, we show the charge assignments of the fields under $U(1)_{B-L}$, and $Z'_2$. 
\begin{table}[htb]
\begin{center}
\begin{tabular}{|c|c|c|c|c|c|c|c|}
\hline
Fields &    $q_{i}$ & $\ell_i$ &$h^+, k^{++}$ & $\varphi$ & $\eta$ & $N_{R_i}$ & $\psi_i$ \\
\hline\hline
$B-L$ &    $1/3$ & $-1$                 &  $2$         & $2$  & $0$&   $-1$ & $1/3,7/3,13/3$ \\
\hline
$Z_2$ &    $+$ & $+$                 &  $+$                                & $+$  & $+$ &  $-$ & $\pm$ \\
\hline
\end{tabular}
\end{center}
\caption{The assignment of $B-L$ charges ($i=1,2,3$).}
\label{tab:B-L}
\end{table}
The Lagrangian for the model can be written as~\cite{Zee-Babu}

\bea
{\cal L}&=& {\cal L}_{\rm SM}+{\cal L}_{\rm Zee-Babu} +{\cal L}_{\rm kin}+ {\cal L}_\Psi -V, \nl
{\cal L}_{\rm Zee-Babu} &=& f_{ab} l_{aL}^{Ti} C l_{bL}^{j}  \epsilon_{ij} h^+
+ h^\prime_{ab} l_{aR}^{T} C l_{bR}^{j}  k^{++} + {h.c}, \nl
{\cal L}_{N_R} &=&  \ol{N_{R_i}}  i \ga^\mu D_\mu N_{R_i}  -{1 \over 2} \Big(\la_{N_{ij}} \varphi \ol{N^c_{R_i}} N_{R_j} + h.c.\Big)\nl
{\cal L}_\Psi &=& \ol{\psi_i}  i \ga^\mu D_\mu \psi_i- m_{\psi_i} \ol{\psi_i} \psi_i 
-f_{12} \Big(\ol{\psi_1} \psi_2 \varphi^* + \ol{\psi_2} \psi_1 \varphi \Big)
-f_{23} \Big(\ol{\psi_2} \psi_3 \varphi^* + \ol{\psi_3} \psi_2 \varphi \Big) \nl
&& -\eta( y_1 \ol{\psi_1} \psi_1 + y_2 \ol{\psi_2} \psi_2 + y_3  \ol{\psi_3} \psi_3), \nl
{\cal L}_{\rm kin} &=& |{\cal D}_\mu h^+|^2 +|{\cal D}_\mu k^{++}|^2 +|{\cal D}_\mu \varphi|^2
{ +{1 \over 2} \left(\partial_\mu \eta\right)^2
+\sum_{i=1}^{3} \ol{\psi_i} i \gamma^\mu {\cal D}_\mu \psi_i}
 -{1 \over 4 } \hat{Z}^\prime_{\mu\nu} \hat{Z}^{\prime \mu\nu}
-{\sin\chi \over 2 } \hat{Z}'_{\mu\nu} \hat{B}^{\mu\nu},
\label{eq:Lag}
\eea
where $V$ is scalar potential, ${\cal D}_\mu =\partial_\mu + i \hat{e} Q \hat{A}_\mu + i \hat{g}_{Z'} Q' \hat{Z}'_\mu$,
and $\hat{B}_{\mu\nu}$, $\hat{Z}'_{\mu\nu}$ are the field strength tensors of $U(1)_Y$, $U(1)'_{B-L}$ gauge field,
respectively.  

We should get non-trivial mixing in $\psi_i$ states to generate TMDO.
After $U(1)_{B-L}$ symmetry breaking,  the mass terms of the Dirac dark fermions are given by
\bea
{\cal L}_{\text{$\psi$ mass}} &=& -
\left(\begin{array}{ccc}\ol{\psi_1} & \ol{\psi_2}  & \ol{\psi_3} \end{array}\right)
\left(\begin{array}{ccc}
m_{\psi_1} & \frac{f_{12} v_\varphi}{\sqrt{2}} & 0\\
\frac{f_{12} v_\varphi}{\sqrt{2}}  & m_{\psi_2} & \frac{f_{23} v_\varphi}{\sqrt{2}} \\
0 &\frac{f_{23} v_\varphi}{\sqrt{2}}   & m_{\psi_3}  
\end{array}
\right)
\left(\begin{array}{c}
{\psi_1} \\
{\psi_2} \\
{\psi_3}
\end{array}
\right),
\eea
which provides the required mixing.
The lightest $\psi_1^\prime$ is absolutely stable due to the local $Z_6$ symmetry and become a DM candidate.  
{ We take $m_{\psi'_1} \sim {\cal O}(1) \, {\rm TeV}$, because this gives not only the correct relic density, but
the necessary self-scattering cross section to solve the small scale structure problems of the CDM,
when the coupling of DM with the light scalar is of order one.}
The $\psi'_2$ can decay into $\psi_1^\prime$ and a photon through the TMDO, which can explain the 3.5 keV X-ray line signal.
It can also be a DM component if its lifetime is much longer than the age of the universe.
To get 3.5 keV X-ray line in the decay process $\psi_2^\prime \to \psi_1^\prime \gamma$ we fix the mass difference
\bea
 \Delta m_{21} \equiv m_{\psi_2^\prime} - m_{\psi_1^\prime} 
= \frac{2 m_{\psi_2^\prime} E_\gamma}{m_{\psi_2^\prime} + m_{\psi_1^\prime}} \simeq E_\gamma =3.5\, {\rm keV},
\eea
where we assumed $m_{\psi_i^\prime} \gg 3.5\, {\rm keV}$.

The effective operator for magnetic transition $\psi'_2 \to \psi'_1 \gamma$, is given by
\bea
{\cal L}_{\rm eff} &=& {1 \over \Lambda} \overline{\psi'_1} \sigma_{\mu\nu} \psi'_2 F^{\mu\nu},
\label{eq:MDO}
\eea
It is generated by so-called ``Barr-Zee'' type two-loop diagrams~\cite{Arhrib:2001xx}.
The state $\psi'_2$ decays almost
100\% via (\ref{eq:MDO}).
Given that $\chi$ and $h-\phi(n)$~\cite{Baek:2011aa} mixing are strongly constrained and the Barr-Zee type diagrams
are generated even in the limits where those mixings vanish, we can consider the effects of non-vanishing mixings
as small perturbations. 
The leading contribution of two-loop Barr-Zee type diagrams to $1/\Lambda$ is obtained to
be
\bea
{1 \over \Lambda}&\simeq&-\sum_{s=h^+,k^{++}}\frac{8 e g^2_{Z'}  \Delta Q_\psi   Q_s Q_{s}^\prime  \lambda_{\varphi s}
\delta^2 \cos 2\theta_{12} s_{13} s_{23} }{(4\pi)^4}  \nl
&\times&\int_0^1 dx \int [d\be]
\frac{x \be_4^2 m_{\psi'_3}^3}
{\left(\be_1 m_{Z'}^2 + \be_2 m_\phi^2 +\be_3 m_s^2/(x(1-x)) +\be_4^2 m_{\psi'_1}^2\right)^2},
\label{eq:d_M}
\eea
where $[d\be] \equiv d\be_1 d\be_2 d\be_3 d\be_4 \delta(1-\be_1 -\be_2 -\be_3 -\be_4)$, 
$\Delta Q_\psi = Q_{\psi'_3}-Q_{\psi'_2}= Q_{\psi'_2}-Q_{\psi'_1}=2$, $\delta=\Delta m_{31}/m_{\psi'_3}$ 
and we neglected small contribution proportional to $\Delta m_{21} (\simeq 3.5 {\rm keV})$.

\begin{figure}
\includegraphics[width=5cm]{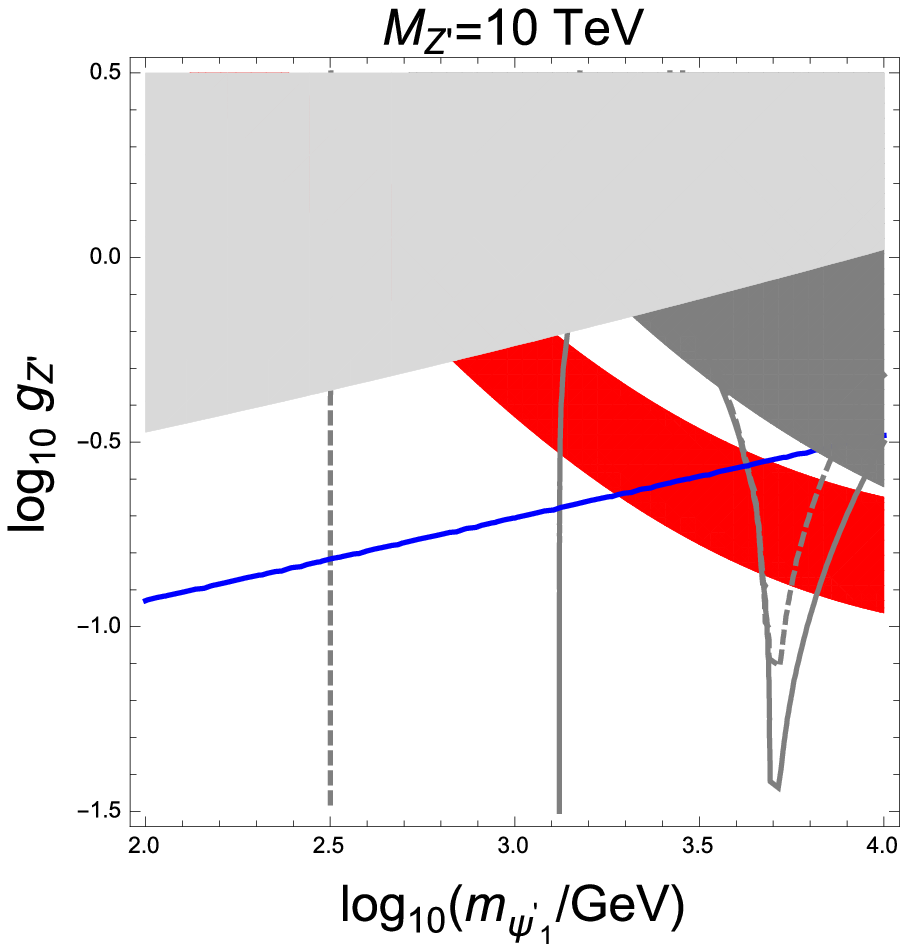}
\includegraphics[width=5cm]{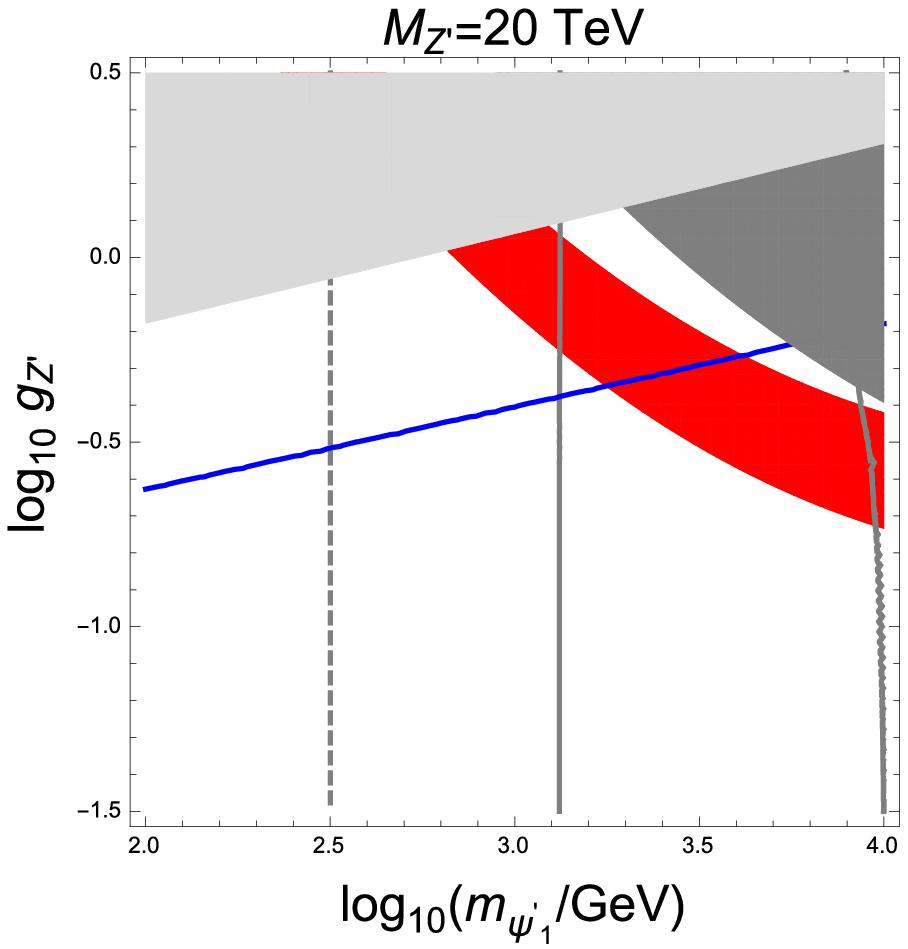}
\caption{Plots in $(m_{\psi'_1},g_{Z'})$-plane. The red-colored region can explain the 3.5 keV X-ray line signal. The
dark gray region is excluded because the lifetime of $\psi'_2$ is shorter than the age of the universe.
The light gray region is excluded by LUX DM direct detection experiment. The blue line is the sensitivity the next
XENONO1T experiment can reach. The black solid (dashed) line gives the correct relic abundance of DM for $y_i=2(1)$. 
For the left (right) plot we set $M_{Z'}=10 (20)$ TeV.}
\label{fig:X-ray}
\end{figure}

In Fig.~\ref{fig:X-ray}, the red-colored region explains the observed
X-ray line signal in the $(m_{\psi'_1},g_{Z'})$-plane. For the left (right) panel we have taken $M_{Z'}=10 \,(20)$ TeV.
For other parameters we have fixed $\delta=0.2$, $\th_{12}=\th_{23} =0.2$, 
$m_\phi=  m_{h^+}= m_{k^{++}}=1$ TeV, $\lambda_{\varphi h} =\lambda_{\varphi k} =1$.
We have checked that the signal region is not very sensitive to the mass parameters,
$m_\phi, m_{h^+}$ and $m_{k^{++}}$.
The black solid (dashed) lines satisfy the observed relic abundance of dark matters,
$\Omega_\psi h^2 =0.1199 \pm 0.0027$, for $y_i=2 \, (1)$.
The vertical lines come from the annihilation channel $\psi_{1(2)} \overline{\psi_{1(2)}} \to n n$ and therefore
are sensitive to the Yukawa couplings $y_i$. There are also resonance regions when $m_{\psi'_1} \approx M_{Z'}/2$.
The region with dark gray color is excluded because it does not satisfy the longevity of the decaying DM. 
The light grey region is excluded by LUX DM direct search experiment and blue line show
the sensitivity of future DM experiment XENON1T.
In our case the direct detection of DM is dominated by $Z'$ boson exchange diagram even though $Z'$ is very heavy,
$M_{Z'}=10 (20) \, {\rm TeV}$.
{We note that $m_{h^+}= m_{k^{++}}=1$ TeV can easily evade the constraints from the lepton flavor violating
processes with $f_{ab} , h'_{ab} \sim {\cal O}(0.01)$, while 
still being able to explain neutrino masses.
}

For light $\eta$ particle ($m_\eta \sim 1-10 \, {\rm MeV}$), $\eta$-exchanging (in)elastic self-interacting processes
$\psi'_{1(2)},\psi'_{1(2)} \to \psi'_{1(2)},\psi'_{1(2)}$ can be strong. When they have cross sections
\bea
 \sigma_T/m_{\psi'_1} \sim 0.1 - 10 \, {\rm cm^2/g},
\eea
we can solve the small scale structure problems such as core-vs-cusp problem and too-big-to-fail problem in our model.
In our model, we need relatively large ($y_i \sim {\cal O}(1)$) Yukawa coupling of $\eta$ with the DM, to get the correct
relic density. 
We can see that the DM scattering cross section can be in the $0.1-10 \, {\rm cm^2/g}$  range for $m_{\psi'_1}=0.1 - 10$ TeV
and $m_\eta=0.1-10$ MeV.

\section{Conclusions\label{sec:Conclusions}}
We extended the Zee-Babu model for neutrino masses to have $U(1)_{B-L}$  gauge symmetry and 
to incorporate Dirac dark matters to explain the X-ray line signal. 
We also introduced $U(1)_{B-L}$ breaking scalar, singlet scalar,
and right-handed neutrinos.
The charges of the particle content are assigned 
in such a way that after the $U(1)_{B-L}$ breaking scalar getting VEV
the local $U(1)_{B-L}$ symmetry is broken down to a discrete symmetry. The lightest Dirac dark fermion $\psi'_1$
whose mass is TeV scale
transforms non-trivially under this discrete symmetry and becomes stable.

{
The heavier $\psi'_2$ particle can decay almost 100\% through the magnetic dipole transition operator
$\overline{\psi'_1} \si_{\mu\nu} \psi'_2 F^{\mu\nu}/\Lambda$.
Since this operator is generated at two-loop so-called Barr-Zee diagrams, the cut-off scale
$\Lambda$ is very high $\sim 10^{15}$ GeV and the lifetime of $\psi'_2$ is much longer than the age of the universe.
And $\psi'_2$ can be a decaying dark matter candidate.
If $\De m_{21} =m_{\psi'_2}-m_{\psi'_1} \simeq 3.5$ keV,
the recently claimed X-ray line signal~\cite{Xray_exp} can be accommodated for wide range of dark matter masses.
}

The relic abundance of dark matters in the current universe can also be explained 
by the dark matter annihilation into two singlet scalars and also by the $Z'$-resonance annihilation.
Although our $Z'$ is very heavy $\gtrsim 10$ TeV, it can still mediate the dark matter scattering off atomic
nuclei at the level that can be probed at the next generation dark matter direct search experiments.
The singlet scalar can be very light  ($m_\eta =0.1-10$ MeV) and mediate strong self-interactions of dark matters
with cross section $\sigma_T = 0.1 -10 \, {\rm cm^2/g}$,
which can solve small scale structure problems, such as the core-vs-cusp problem and the too-big-to-fail problems,
of the standard $\Lambda$CDM model.

{ The small mass difference $\Delta m_{21}$ and the small VEV of $\eta$ are technically natural in the sense
of 't Hooft. The singlet scalar and the right-handed neutrinos decay fast without causing any cosmological problems.}

\begin{acknowledgments}
 This work was supported by in part NRF Grant 2012R1A2A1A01006053.
\end{acknowledgments}

\bigskip 

\end{document}